\documentclass[
 reprint,
 amsmath,amssymb,
 aps,
]{revtex4-2}

\usepackage{graphicx}
\usepackage{dcolumn}
\usepackage{bm}

\begin{document}

\title{Extraction of geometric and transport parameters from the time constant of exocytosis transients measured by nanoscale electrodes}

\author{Sundeep Kapila}
\email{sundeep@ee.iitb.ac.in}
\author{Pradeep R. Nair}
\email{prnair@ee.iitb.ac.in}
\affiliation{Department of Electrical Engineering, Indian Institute of Technology Bombay, Mumbai, India}
\date{\today}


\begin{abstract}
Exocytosis is a fundamental process related to the information exchange in the nervous and endocrine system. Among the various techniques, vesicle impact electrochemical cytometry (VIEC) has emerged as an effective method to mimic the exocytosis process and measure dynamic information about content transfer using nanoscale electrodes. In this manuscript, through analytical models and large scale simulations, we develop scaling laws for the decay time constant $(\tau)$ for VIEC single-exponential transients. Specifically, our results anticipate a power law dependence of $\tau$ on the geometric and the transport parameters. This model compares very well with large scale simulations exploring the parameter space relevant for VIEC and with experimental results from literature. Remarkably, such physics based compact models could allow for novel multi-feature based self consistent strategies for back extraction of geometric and transport parameters and hence could contribute towards better statistical analysis and understanding of exocytosis transients and events. 
\end{abstract}

\maketitle

\section{\label{sec:intro}Introduction}
Information exchange within multicellular organisms occur through the release of specific bio-molecules (e.g., hormones, neurotransmitters) into body fluid like blood or a restricted volume like synaptic gap between two neurons \cite{intro1,singlecell,unified}. This release of bio-molecules happens through exocytosis, a process of fundamental importance, where the bio-molecules are carried in a vesicle within the cell, then the vesicle fuses with the membrane and the content is released through a pore opening in the cell membrane. Extensive research, both experimental \cite{review_elec_neuro,amp_recent,exp_amp_recent,exp_elec_imaging} and numerical \cite{analytical1,analytical2}, has been reported to analyze the time dynamics of exocytosis and hence decipher the underlying physical processes (like pore opening, closure, kiss and run, etc.)\cite{kiss_and_run,kiss_and_run2,partial_release,fullfusion,actincontrol}. Experimental approaches like electrochemical detection of released bio-molecules \cite{nano1,mos,biovesicles} and optical methods like super-resolution stimulated emission depletion (STED) microscopy \cite{sted}, total internal reflection fluorescence microscopy (TIRFM) \cite{dual_method}, and numerical approaches are routinely used to gather insights on the time dynamics of exocytosis and hence extract geometric and transport parameters \cite{amatore-reverse}. In addition, literature also reports the impact of factors like the electrode geometry \cite{electrode,nano_conical}, pH variation \cite{ph_factor}, and pore dynamics \cite{pore_exp} on the exocytosis signal.

Recent experimental techniques like vesicle impact electrochemical cytometry (VIEC) \cite{viec-first} have emerged as a useful scheme to gather more information on the vesicle geometry, its contents and the biological processes driving the release of the vesicular content by enabling detection and characterisation of single-vesicle electrochemical events \cite{single-vescile}. In this technique, a polarised ultramicroelectrode (UME) is placed in a suspension of single vesicles, which rupture on impact with the electrode and release its contents. These contents are detected by the electrode being in close proximity ($\approx 20$ nm) of the vesicle and measured electrochemically as a current signal. This technique mimics the exocytosis process except that in VIEC the vesicle attaches to the electrode through electroporation and the vesicle pore remains open. This triggers the release of the full contents of the vesicle, which does not happen in all exocytosis events. However, certain kind of exocytosis events have been reported where the pore remains static after opening and releases its contents \cite{exocytosis_review,exo_endo_review} The transient decay signal of VIEC will be a good representative of such exocytosis events. The advantage of VIEC is that it gathers information regarding single-vesicle events and hence is able to provide with insights into the specific biological processes or factors \cite{viec-biop} that can cause variation in the various output parameters of the exocytosis process - like maximum current ($I_{max}$), the half time ($t_{\frac{1}{2}}$), the rise time ($t_{rise}$), the fall time ($t_{fall}$), the decay time constant ($\tau$), and the quantal or total molecules released ($Q$) \cite{viec-cnp}. By combining VIEC with techniques like resistance pulse \cite{biovesicles,viec-rp} or carbon nano-pipettes \cite{viec-cnp}, literature also reports estimation and correlation of geometric parameters like the vesicle radius and the vesicle concentration \cite{viec-rp}. 

While the above developments are encouraging, the time constants associated with VIEC transients are yet to be characterized in detail and there are several open questions. For example, what would be the dominant parameters that influence the time constants? What information can be obtained from the time constants regarding the size of vesicles and transport properties? Evidently, such detailed information, along with the existing characterization schemes, could allow better estimation of the geometric parameters like the pore width and the transport parameters like the diffusion coefficient within the vesicle. These estimates can then be used to understand the various biological processes involved in static pore exocytosis events.

In view of the above, here we develop an analytical model  for the time constants associated with VIEC transients. The model predictions are then calibrated against detailed numerical simulations to develop a closed form expression for $\tau$. The predictions from the closed form expression are found to be inline with reported experimental data and also have the potential to improve the estimates of geometric and transport parameters by eliminating the need for some assumptions in the current estimation techniques. Below, we first describe the model system to explore VIEC transients. 

\section{Model System}

\begin{figure}[t!]
\centering
\includegraphics[width=\columnwidth]{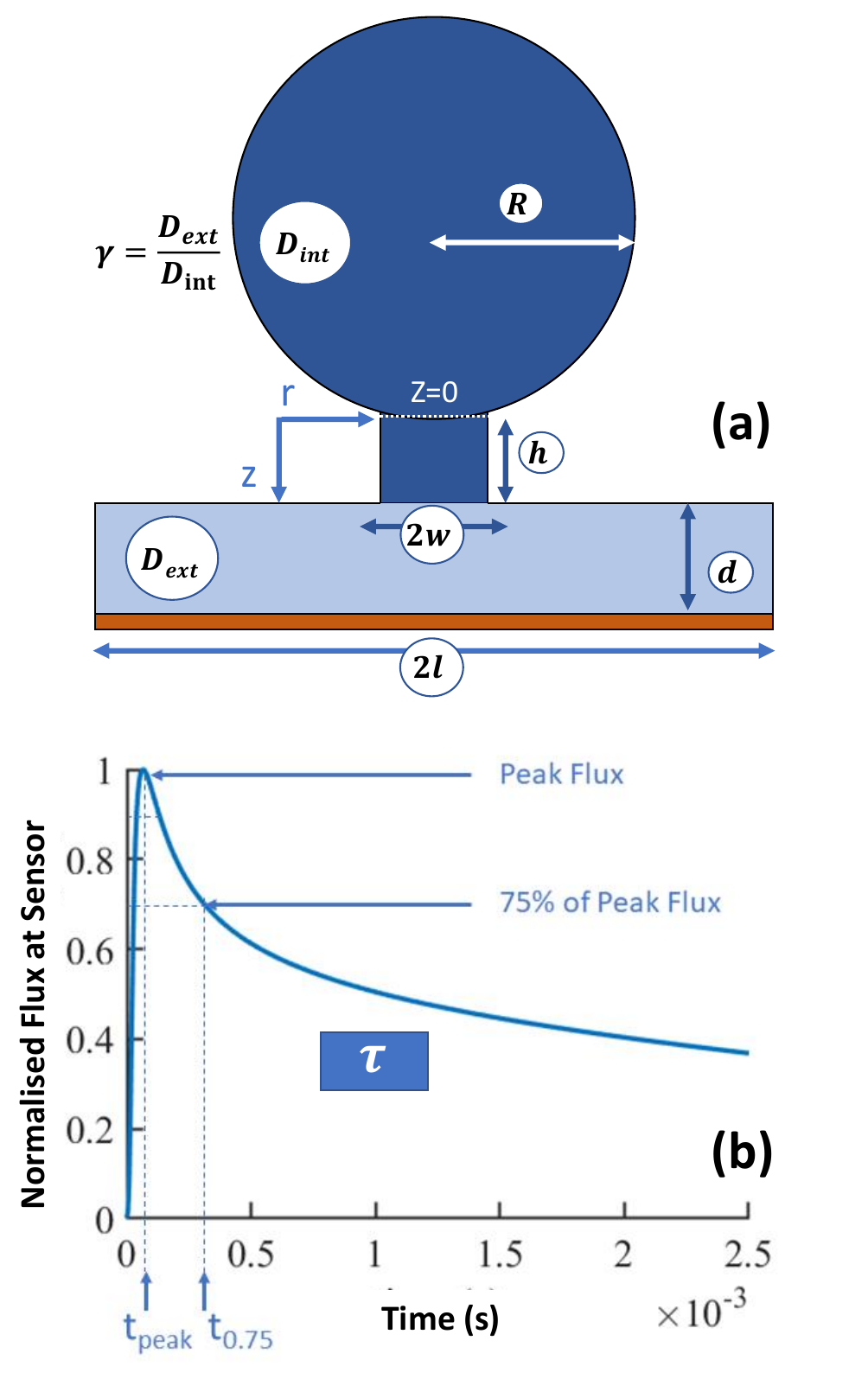}
\caption{\label{fig:model_output} Model System. (a) VIEC system layout and key parameters: vesicle radius $(r)$, pore width $(w)$, pore height $(h)$, sensor distance $(d)$, and sensor radius $(l)$. (b) Typical transients measured at sensor. The monotonic decay from $t_{0.75}$ is usually characterised as a single exponential with time constant $\tau$.}
\end{figure}

In VIEC, vesicles filled with bio-molecules come in contact with an UME resulting in the opening of a pore (see Fig. \ref{fig:model_output}a). As the pore opens at time $t=0$, the  bio-molecules diffuse through the pore and reach the sensor. Electrochemical reaction at the UME results in a current (see Fig. \ref{fig:model_output}b) which is proportional to the net flux of bio-molecules at the sensor surface. Evidently, as the pore opens, the current increases sharply  due to the sudden  influx of bio-molecules. With time the concentration of molecules in the extra-vesicular medium increases and this reduces the concentration gradient across the pore and hence the diffusive flux of bio-molecules from the vesicle decreases. As a result, the sensor current exhibits a decreasing trend after reaching a peak. The initial part of the decay (till signal reaches $75\%$ of the peak) is a combination of the pore width opening and diffusion through the pore and extra-vesicular space \cite{analytical1,analytical2,amatore-reverse}. The decay of the transient post $75\%$ of the peak can be characterised as a single exponential decay (see Fig. \ref{fig:model_output}b) with a time constant $(\tau)$ \cite{biovesicles}, especially in the case of non-dense core vesicles, which is the scenario we are considering in this manuscript.  Accordingly, the key parameter that characterises the measured signal in non-dense core vesicles is the time constant ($\tau$) of the exponential decay of the signal post $75\%$ of peak.

In this manuscript, we aim to unravel the functional dependence of the long term decay constant, namely $\tau$ on the geometric and transport parameters mentioned earlier (listed in Fig. \ref{fig:model_output}). This could enable reconstruction of the transport processes and geometry from statistical analysis of the decay transient signals. To this end, transport of bio-molecules through the pore and final detection at the sensor needs to be modeled using transient diffusion equation. As shown in Fig. \ref{fig:model_output}a, our model system consists of a spherical vesicle attached to a membrane with a cylindrical pore and a large planar disc shaped sensor at a distance $d$ from the membrane. The corresponding transient diffusion equation, using the cylindrical symmetry of the system, is given as

\begin{equation} \label{eq:diff_eq2}
\frac{\delta C(r,z,t)}{\delta t} = \vec{\nabla} . \left[ D(r,z) \vec{\nabla}{C}(r,z,t) \right]
\end{equation}

where $C$ is the concentration and $D(r,z)$ is the position dependent diffusion coefficient of bio-molecules. The axes $r$ and $z$ are as shown in Fig. \ref{fig:model_output}a. The diffusion process could be influenced by the presence of other biological structures and proteins like microtubules, actin and membrane proteins \cite{actincontrol}. Accordingly, the diffusion coefficients  are assumed to be different inside and outside the vesicle \cite{analytical1}. We assume the vesicle is initially uniformly filled with a concentration $C_v$. Accordingly, geometric parameters that define the time dynamics of the release event are vesicle radius $(R)$, pore width $(w)$, pore height $(h)$, and sensor distance $(d)$, while the relevant transport parameters are the diffusion coefficient ($D_{ext}$) outside the vesicle (marked in light blue color), the diffusion coefficient($D_{int}$) inside the vesicle (marked in darker blue color) and their ratio $(\gamma)$ (see Fig. \ref{fig:model_output}a.) 

\section{Predictive models for exocytosis time constant}

The time dynamics of exocytosis could be influenced by several phenomena like pore opening, diffusion of bio-molecules, pore closing, etc. However, a recent study indicated that the pore width remains unchanged in a significant fraction of exocytosis events \cite{sted}. In VIEC, pore opens to a maximum width and then remains unchanged \cite{biovesicles}. Accordingly, here we explore the time dynamics of an event where the pore width remains unchanged after it reaches its peak. Based on Eq. \ref{eq:diff_eq2}, we first develop an analytical model to predict the functional dependence of the time constant on the vesicle and pore geometry and the transport parameters. Later, these analytical  predictions are compared against detailed numerical simulations and reported experimental data.

The time constants for the exponential decay of VIEC transients can be estimated once we identify an appropriate rate equation for the flux at the sensor surface. To this end, ignoring the variation along the radial direction in the pore, let $C(z,t)$ denotes concentration at time $t$, at co-ordinate $z$ along the $z$ direction with $z=0$ at the pore and vesicle interface and $z=h$ at the pore and external interface (see Fig. \ref{fig:model_output}a). The time constants associated with $C(0,t)$ can be obtained as follows:

In a first order analysis, conservation of total number of particles in the vesicle leads to
\begin{equation}
    C(0,t) = C(0,0) - \frac{3}{4 \pi R^3} \int_{0}^t Flux(0,t)dt
\end{equation}
Differentiating the above equation, we get
\begin{equation} \label{eq:conservation}
    \frac{\delta C(0,t)}{\delta t} = -\frac{3}{4 \pi R^3}Flux(0,t)
\end{equation}
where $Flux(0,t)$ is the time dependent total flux through the pore at $z=0$. The diffusion happens through a distance of $h$ through the pore and then through the extra-vesicular space to the sensor surface (where $C=0$ as the boundary condition). Hence, $Flux(0,t)$ can be represented in terms of effective distance  $h_{eff}$ as
\begin{equation} \label{eq:flux_time}
    Flux(0,t) \approx  \pi w^2 D_{int} \frac{C(0,t)}{h_{eff}}
\end{equation}
Substituting the expression for $Flux(0,t)$ from equation \ref{eq:flux_time} in equation \ref{eq:conservation}, we get

\begin{equation} \label{eq:tau_eqn}
    \frac{\delta C(0,t)}{\delta t} = -\frac{3 w^2 D_{int}}{4R^3 h_{eff}} C(0,t) 
\end{equation}

The above is a linear differential equation with a time constant. Since the $Flux(0,t) \propto C(0,t)$, the rate equation for $Flux(0,t)$ also has the same form as given by \ref{eq:tau_eqn}. 
Once the pore has opened completely, the flux at the sensor surface is expected to be a scaled and a delayed version of $Flux(0,t)$ (also influenced by the respective diffusion coefficients). Accordingly, we expect the time constants of the flux at the sensor surface (decay constant of the transient in VIEC setup) to also retain the same functional dependence with key geometric parameters like $R$, $w$, $h$, and $d$ and transport parameters namely $D_{ext}$ and $D_{int}$. Hence, by retaining the power law dependence on the geometric and transport parameters, the time constant for the VIEC transients measured at the sensor is given as:
\begin{equation} \label{eq:tau_final}
    \tau = K\frac{R^3 h_{eff}}{w^2 D_{int}}
\end{equation}
where $K$ is a dimensionless constant. From this we can observe that $\tau \propto R^3$, $\tau \propto w^{-2}$, and $\tau$ is dependent only on the intra-vesicular diffusion and is independent of the extra-vesicular diffusion. The terms $h_{eff}$ and the constant $K$ will be evaluated through simulations, as described in the next section. 

\section{Simulation Results and Analysis}

\begin{figure}[ht]
\centering
\includegraphics[width=0.9\columnwidth]{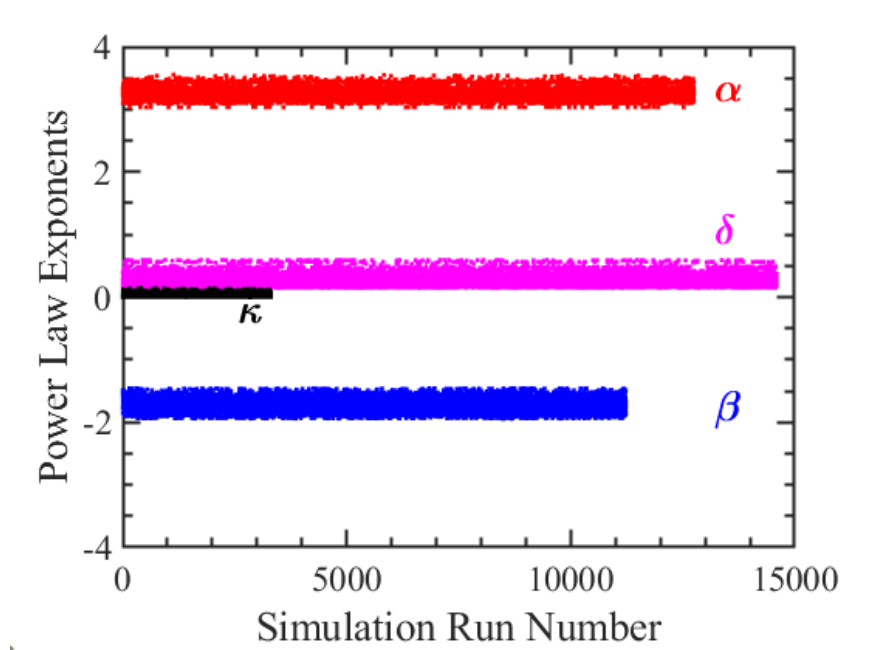}
\caption{\label{fig:power_all3}Analysis of numerical simulations to identify power law exponents involved in $\tau \propto
R^\alpha w^\beta h^\delta d^\kappa $. The methodology used to obtain the exponents is described in the text. 
}
\end{figure}

\indent To test the analytical predictions, transient diffusion equation was numerically solved for the model system with cylindrical symmetry (see Eq. \ref{eq:diff_eq2}). Finite difference scheme with BDF2 time integration was employed for the numerical solution of the discretized equations. Similar approach was earlier used, for different geometries, to address transient diffusion towards regular and fractal surfaces \cite{prnair1,prnair2,prnair3,prnair4,prnair5}. Extensive simulations were performed to explore the influence of geometry and transport by varying the following five parameters: (i) vesicle radius $(R)$ from $45$ to $100$ nm in steps of $5$ nm, (ii) pore width $(w)$ from $10$ to $75$ nm in steps of $5$ nm and for condition where $w<0.8R$, (iii) pore height $(h)$ from $5$ to $10$ nm in steps of $1$ nm, (iv) sensor distance $(d)$ from $10$ to $35$ nm in steps of $1$ nm, and (v) The ratio of the diffusion coefficients ($\gamma$) was varied from 5 to 100 to keep it consistent with the scenario of $D_{ext} >> D_{int}$ \cite{gamma_measurement}. We kept the radius $(l)$ of the disk shaped sensor fixed at $200$ nm to model the case of sensor being much larger than the pore width, pore height and sensor distance, i.e. $l >> w,h,$ and $d$. The diffusion coefficient in the intra-vesicular space $(D_{int})$ was assumed to be $3.5 \times 10^{-8}$ cm$^{2}$/s. These values, especially the sensor distance was chosen to focus on techniques like vesicle impact electrochemical cytometry (VIEC) \cite{biovesicles} and measurements where the sensor is brought in to touch with the vesicle or cell boundary \cite{pore_exp}.\\
\indent We ran $\approx 85,000$ individual simulations (each for a time duration of $5$ ms) to explore the above mentioned parameter space. We computed the time constant $(\tau)$ for every combination of the above mentioned geometric and transport parameters. The transients with $\tau \le 10$ ms are considered for further statistical analysis   to extract the functional dependencies on the key geometrical parameters. Interestingly, our analysis finds that $\tau$ varies as a power-law with relevant parameters - i.e.,   $\tau \propto R^\alpha w^\beta h^\delta d^\kappa $.  Such numerically extracted $\alpha$, $\beta$, $\delta$, and $\kappa$  are plotted against various simulation runs in Fig. \ref{fig:power_all3} for all the six different values of $\gamma = D_{ext}/D_{int}$, i.e. $\gamma$ in $ \left[ 5,10,25,50,75,100 \right]$. To extract the exponent for a particular parameter, the other four parameters were kept constant. In this way, multiple values of the exponent were extracted from the simulations. For example, to extract $\alpha$, unique combinations of the other four parameters, namely $\gamma, w, h$, and $d$  are considered. We have considered 12,753 such unique combinations of $(\gamma, w, h$, $d)$, which are represented on the x-axis as 'Simulation Run Number' in Fig \ref{fig:power_all3}. For each such combination, the variation of $\tau$ as a function of $R$ is considered and the corresponding  $\alpha$ is extracted as the slope of the best fit line  of $log(\tau(R))$ vs $log(R)$. Such extracted $\alpha$ is plotted on the y-axis as the 'Power Law Exponent' in Fig \ref{fig:power_all3}. Similar method is used to plot the other power law exponents shown in Fig \ref{fig:power_all3}. 

The simulation results indicate that $\tau$ varies as $\approx {R^3}$ and $\approx w^{-2}$, which is close to the power law anticipated by our analytical model (see equation \ref{eq:tau_final}). The figure also shows that the variation in the value of $\alpha$ and $\beta$ is in a very narrow range indicating that the power law is a good estimate for the decay constant. Table \ref{table:power_law_summary_all} summarises the average value (Mean) and standard deviation (SD) of $\alpha$ and $\beta$ for the different values of $\gamma$. From the table \ref{table:power_law_summary_all}, we observe that the average value across all $\gamma$ for  $\alpha$ is $3.25$ and for $\beta$ is $-1.73$. While deriving equation \ref{eq:tau_final}, it was assumed that the concentration at the sensor would have the same time constant as the concentration at the vesicle and pore interface. The variation in the average value of $\alpha$ and $\beta$ are possibly related to the impact of the extra-vesicular diffusion on the time constant of the flux at the sensor. This is also captured in the parameters $h_{eff}$ and $K$ from equation \ref{eq:tau_final}. We also evaluated the parameters $\delta$ and $\kappa$ as power law exponents for $h$ and $d$ respectively. The plots are shown in Fig \ref{fig:power_all3} and the data is reproduced in the Table \ref{table:power_law_summary_all}.

\begin{table}[ht]
\caption{Power law exponent of $\tau \propto
R^\alpha w^\beta h^\delta d^\kappa $. Mean $\pm$ SD for $\alpha$, $\beta$, $\delta$, and $\kappa$ }
\label{table:power_law_summary_all}
\centering
\begin{ruledtabular}
\begin{tabular} {|l|c|c|c|c|}
\hline
$\gamma$ & $\alpha$ & $\beta$ & $\delta$ & $\kappa$ \\
\hline
5   & 3.31 $\pm$ 0.09 & -1.77 $\pm$ 0.08 & 0.25 $\pm$ 0.07 & 0.075 $\pm$ 0.025 \\
10  & 3.27 $\pm$ 0.08 & -1.74 $\pm$ 0.09 & 0.26 $\pm$ 0.08 & 0.041 $\pm$ 0.013 \\
25  & 3.24 $\pm$ 0.08 & -1.72 $\pm$ 0.10 & 0.27 $\pm$ 0.09 & 0.016 $\pm$ 0.004 \\
50  & 3.24 $\pm$ 0.08 & -1.71 $\pm$ 0.11 & 0.27 $\pm$ 0.09 & 0.007 $\pm$ 0.002 \\
75  & 3.23 $\pm$ 0.08 & -1.71 $\pm$ 0.11 & 0.27 $\pm$ 0.09 & 0.004 $\pm$ 0.001  \\
100 & 3.23 $\pm$ 0.08 & -1.71 $\pm$ 0.11 & 0.27 $\pm$ 0.09 & 0.003 $\pm$ 0.001 \\
\hline
Avg & 3.25 $\pm$ 0.09 & -1.73 $\pm$ 0.10 & 0.27 $\pm$ 0.09 & 0.024 $\pm$ 0.028\\
\hline
\end{tabular}
\end{ruledtabular}
\end{table}

\begin{figure}[ht]
\centering
\includegraphics[width=0.9\columnwidth]{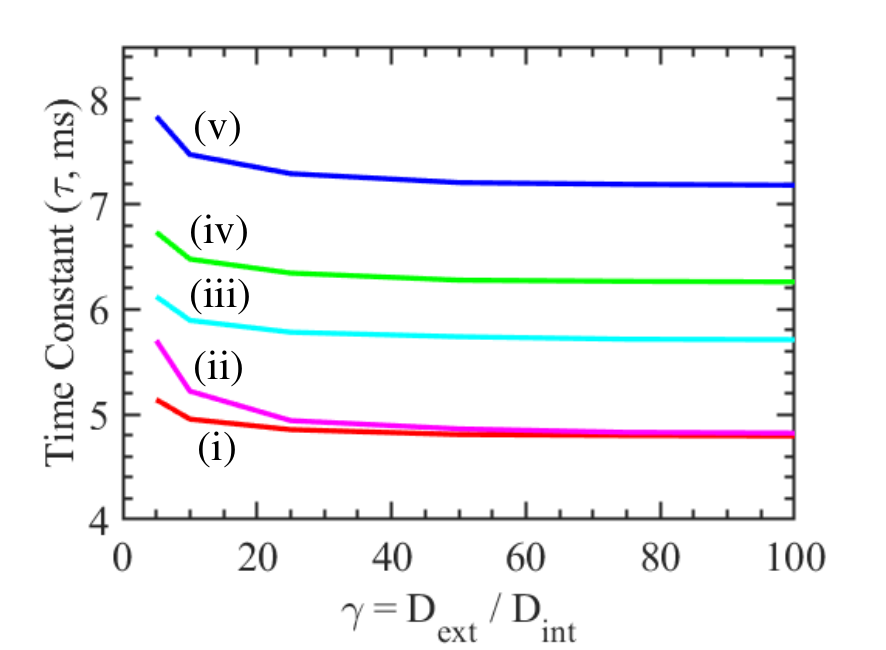}
\caption{\label{fig:dout_zero} Variation of $\tau$  as a function of  $\gamma = D_{ext}/D_{int}$ for different combinations  of geometric parameters. Here, the parameters for base case indicated by (i) are $R = 90$ nm, $w = 40$ nm, $h = 5$ nm, $d = 10$ nm. In each of the other cases only one parameter is modified, in comparison to the base case, as follows (ii) $d = 30$ nm, (iii) $h = 10$ nm, (iv) $R = 100$ nm, (v) $w = 30$ nm.}
\end{figure}

We further explored the impact of the extra-vesicular diffusion co-efficient on the decay constant. When $D_{ext} >> D_{int}$,   the entire process will be limited by the diffusion within the pore as observed in \cite{analytical1}. Hence, the impact of the extra-vesicular diffusion coefficient $(D_{ext})$ on the decay constant will reduce as $D_{ext}$ increases . This is confirmed by Fig \ref{fig:dout_zero} which indicates that for different values of the geometric parameters $(r,w,h,d)$, the decay constant $(\tau)$ reduces with increasing $\gamma$ and then asymptotically becomes a constant for $\gamma \ge 25$.

Table \ref{table:power_law_summary_all} indicates that the average value across all $\gamma$ for $\delta$ is $0.27$ and for $\kappa$ as $0.03$. Based on the average values of $\alpha$, $\beta$, $\delta$, and $\kappa$, the corresponding space dimensions add up to $1.82$ vis-a-vis the expected value of $2$ (see eq. \ref{eq:tau_final}). With these fractional exponents for the key parameters, the resultant constant $K$ will no longer remain dimensionless - which is not a convenient scenario for ease of analysis. Hence, to maintain the dimensionless nature of $K$, we distribute the error of $0.18$ equally to all the four dimensions giving the exponents of $r, w, h, $ and $d$ as $3.30, -1.68, 0.31$ and $0.07$ respectively. Accordingly, based on the insights from the large scale simulations across the parameter space, equation \ref{eq:tau_final} can be updated as 
\begin{equation} \label{eq:tau_eff}
    \tau = K\frac{R^{3.30} h^{0.31}d^{0.07}}{w^{1.68}D_{int}} \left(1+1/\gamma \right)
\end{equation}
where $K$ needs to be evaluated from simulations. Here the factor $(1+1/\gamma)$ accounts for the minor dependence on the diffusion coefficients, as indicated by Fig. \ref{fig:dout_zero}.

\begin{figure}[ht]
\centering
\includegraphics[width=0.9\columnwidth]{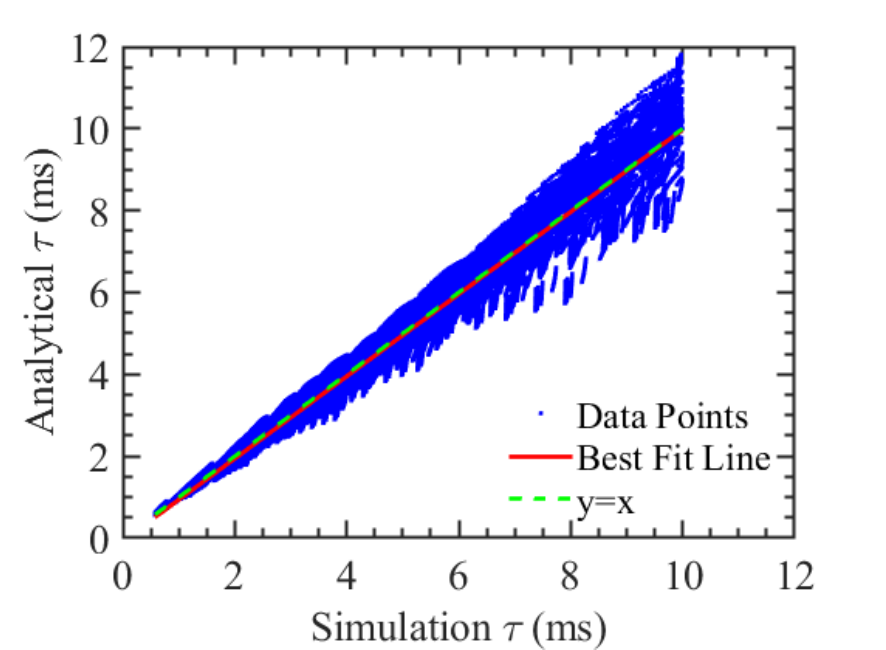}
\caption{\label{fig:scatter_gamma}  Comparison of analytical $\tau$ (eq. \ref{eq:tau_eff} with $K=1.39$) with numerical simulations for the entire parameter space. The best fit line and the $y=x$ line are provided for visual reference.}
\end{figure}

Based on equation \ref{eq:tau_eff}, we calculated the predicted $\tau$ and compared it with the $\tau$ extracted from the simulation for the same geometric and transport parameters. Fig \ref{fig:scatter_gamma} shows the scatter plot of the simulation $\tau$ vs analytical $\tau$ for 85,503 data points. This shows a linear relationship with a co-relation coefficient of $0.9894$ and the slope of the best fit line as $1.0044$. The proportionality constant $K$ from equation \ref{eq:tau_eff} was calculated to match the best fit line and was computed to be $1.39$. The root mean square percent error (RMSPE) of the analytical $\tau$ with simulation $\tau$ is $6.44\%$. 

\begin{figure}[ht]
\centering
\includegraphics[width=0.9\columnwidth]{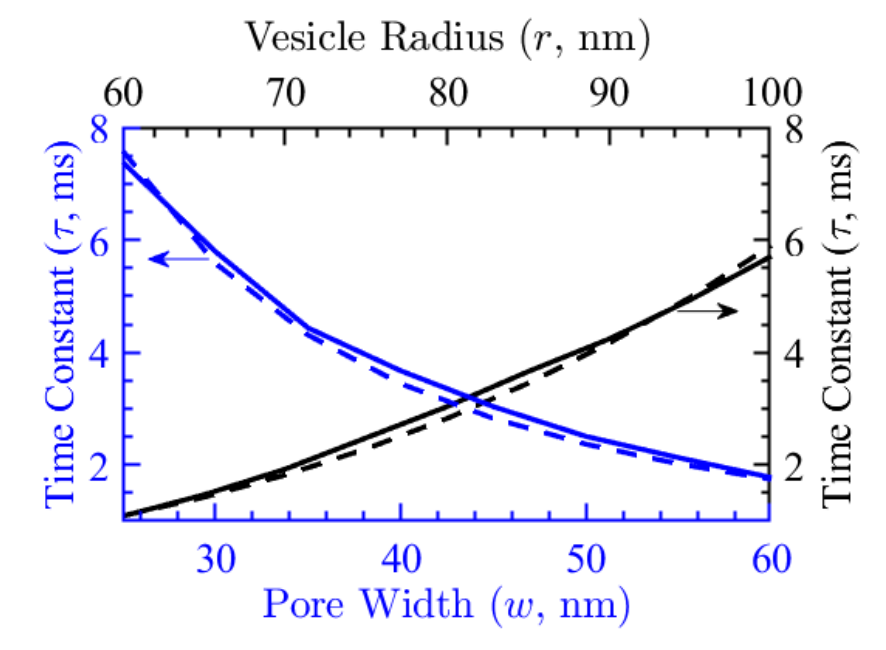}
\caption{\label{fig:match} Comparison of simulation results (solid lines) with $\tau$ from equation \ref{eq:tau_eff} (dashed lines) as function of various geometric parameters.  For both set of results, we have used $\gamma = 10, h = 5$ nm and $d = 20$ nm. The blue curves correspond to $R = 80$ nm while the black data sets correspond to $w = 45$ nm.  }
\end{figure}

\indent Hence, we propose that equation \ref{eq:tau_eff} with   $K=1.39$ is a good expression for $\tau$ in the parameter space relevant for VIEC.  Indeed, Fig \ref{fig:match} shows that the $\tau$ from numerical simulations (in solid lines) are well anticipated by the equation \ref{eq:tau_eff} (in dashed lines) for different combination of geometric and transport parameters ( $\gamma, h, $  and $d$) that are relevant for VIEC applications \cite{biovesicles}. In Fig. \ref{fig:match}, the curves in blue (plotted against the left Y axis) indicate the variation of the time constants as a function of pore width at a specific vesicle radius (80 nm). Similarly, the data sets shown in black (and plotted against the right Y axis) illustrate the dependence of  time constant on vesicle radius (with the pore width being 45 nm). We  observe that the analytical predictions (dashed lines) based on equation \ref{eq:tau_eff} match well with the values extracted from the numerical simulations (solid lines).\\

\section{Application to experimental data}
\indent The key insight that emerges from our analytical predictions and subsequently verified by numerical simulations is the analytical expression for $\tau$ given in equation \ref{eq:tau_eff}. Here,  we compare the model predictions with experimental results reported in \cite{biovesicles}. \\ 
\indent VIEC and resistance pulse (RP) techniques are combined to simultaneously measure vesicle radius and the transient signal for each vesicle \cite{biovesicles,viec-rp}. Further, ref. \cite{biovesicles}  proposed a methodology named finite element simulation algorithm (FESA) to estimate the pore width $(w)$ by matching the experimental results with numerical simulations. FESA is compared with the methodology proposed in \cite{amatore-reverse} and is asserted as better as FESA works for all vesicle radius $(R)$ unlike the methodology proposed in \cite{amatore-reverse}, which requires $R$ to be assumed. While FESA works for all $R$, it requires assumptions on pore height $(h)$, sensor distance $(d)$, $D_{ext}$ and $D_{int}$ to estimate pore width $(w)$. While $h$,$d$, and $D_{ext}$ have good estimates, the real value of $D_{int}$ may vary by each vesicle due to variation in the biological environment (e.g. microtubule network) in each vesicle. Eq \ref{eq:tau_eff} provides a closed form relationship between $D_{int}$ and $w$, Hence, we propose that by  applying FESA self-consistently with eq. \ref{eq:tau_eff}, we can (i) improve the estimate of $w$ and (ii) provide an estimate for $D_{int}$ for the specific vesicle.\\
\indent Our model predictions compare very well with experimental data and allows better estimates for back extraction of key parameters related to excoytosis. For example,reference \cite{biovesicles} has reported vesicle radius $(R)$ measured through RP, and the estimated pore width $(w)$ through FESA. This estimate relies on assumed values for $h,d,D_{ext},$ and $D_{int}$. Based on the above geometric and transport parameter values we computed the $\tau$ based on equation \ref{eq:tau_eff} and compared it with the $\tau$ extracted from the VIEC transients shared in \cite{biovesicles} for five VIEC events. Table \ref{table:exp_summary} summarises the data. From table \ref{table:exp_summary}, we observe that the model predictions compare very well with the experimentally measured $\tau$ - especially, given the fact that $\tau$ depends on at least 5 parameters (see eq. \ref{eq:tau_eff}) whose a-priori and independent estimates are hard to arrive at. \\
\indent There is a significant difference between analytical prediction and experimental result for one case listed as $SN=4$  which could be due to assumption on the value of $D_{int}$ in FESA algorithm. Specifically, table \ref{table:exp_summary} indicates that between data sets 4 and 5, $w$ has decreased while $R$ has increased. Assuming other parameters the same, both of these factors are expected to result in an increase in $\tau$ as the diffusion gets slower. However, we notice the experimental $\tau$ has decreased. This apparently puzzling trend could be the result of the FESA algorithm where the value of $D_{int}$ is assumed to be the same across all vesicles. However, $D_{int}$ could vary across vesicles due to the internal biological structures. \\
\indent It is well known that in such multi-parameter inverse problems, multiple possible solutions for parameter values exist \cite{inverse_problem201511}. Hence, there could be other possible solutions for the tuple $(D_{int},w)$ which might resolve the unexpected scaling trends of $\tau$ with $R$ and $w$. We note  that usage of FESA scheme self-consistently with equation \ref{eq:tau_eff} could yield a better estimate of the tuple $(D_{int},w$). An accurate estimation of pore width $(w)$ would be of significant relevance as \cite{biovesicles} indicates a potential intrinsic relationship between the maximum pore width and the vesicle radius. In addition, a methodology for estimating intra-vesicular diffusion would be significant as it is an indicator of the density of the biological structures inside the vesicle.\\  
\indent In summary, our proposed analytical expression in  equation \ref{eq:tau_eff} for $\tau$ compares well with extensive numerical simulations for VIEC and reported experimental data. With a closed form expression for the functional dependence of $\tau$, our model enables better estimation of pore width and intra-vesicular diffusion coefficient by combining with the FESA methodology in the case of VIEC with resistance pulse. 
 
\begin{table}[ht]
\caption{Comparision of Experimental and Analytical $\tau$ }
\label{table:exp_summary}
\centering
\begin{ruledtabular}
\begin{tabular} {|l|l|c|c|c|c|}
\hline
 \multicolumn{4}{|c|}{}& \multicolumn{2}{c|}{$\tau$ (ms)} \\
\hline

SN &Fig Ref & $w$ (nm) & $R$ (nm)  & Exp & eqn \ref{eq:tau_eff} \\
\hline 
1 & \cite{biovesicles} SI Fig S2  & 138 & 316 & 3.5 & 2.3\\
2 & \cite{biovesicles} SI Fig S2  & 113 & 287 & 3.4 & 2.4\\
3 & \cite{biovesicles} SI Fig S2  & 31  & 208 & 7.0 & 7.2\\
4 & \cite{biovesicles} SI Fig S2  & 45  & 189 & 6.3 & 2.8\\
5 & \cite{biovesicles} Fig 2      & 33  & 220 & 5.6 & 7.8\\

\hline
\end{tabular}
\end{ruledtabular}
\end{table}

\section{Conclusions}
 In this manuscript, we  developed an analytical model for the decay time constant as a function of the geometrical and transport parameters for non-dense core vesicles in a VIEC like setup. The model predictions are well supported by results from the detailed numerical simulations and experimental data reported in the literature. The predicted power law holds true for a wide range of values for the geometric and transport parameters relevant for methods like VIEC where the sensor is kept in close proximity to the vesicle in non-dense core vesicles. We also demonstrated that this power law is valid in the scenario of different diffusion coefficients in the intra-vesicular and extra-vesicular space. These inferences applied self-consistently with the FESA algorithm can improve the accuracy of estimating pore width and also provide a methodology to estimate the intra-vesicular diffusion rate from the experimental data. In future, the model could be updated for scenarios with dense core vesicles and dynamic pore size variation. 

\begin{acknowledgments}
The authors acknowledge and thank Prof. Bhaskaran Muralidharan of the Department of Electrical Engineering, IIT Bombay for insightful discussions. PRN acknowledges Visvesvaraya Young Faculty Fellowship. SK acknowledges Swasth Research Fellowship.
\end{acknowledgments}

\bibliography{predictivemodel}

\end{document}